\documentclass{ws-mpla}

\begin{document}

\markboth{D. Jido, T. Hyodo, A. Hosaka}
{The structure of $N(1535)$ in the aspect of chiral symmetry}

\catchline{}{}{}{}{}

\title{THE STRUCTURE OF $N(1535)$ IN THE ASPECT OF CHIRAL SYMMETRY}

\author{\footnotesize D. JIDO}

\address{Yukawa Institute for Theoretical Physics, Kyoto University, Sakyo, Kyoto, 606-8502, Japan\\
jido@yukawa.kyoto-u.ac.jp}



\author{T. HYODO}

\address{Physik-Department, Technische Universit\"at M\"unchen, D-85747 Gerching, Germany\\
Yukawa Institute for Theoretical Physics, Kyoto University, Sakyo, Kyoto, 606-8502, Japan}

\author{A. HOSAKA}

\address{Research Center for Nuclear Physics, Osaka University, Ibaraki, Osaka 567-0047, Japan }

\maketitle

\pub{Received (Day Month Year)}{Revised (Day Month Year)}

\begin{abstract}
The structure of $N(1535)$ is discussed in dynamical and symmetry aspects based on chiral symmetry. We find that the $N(1535)$ in chiral unitary model has implicitly 
some components other than meson-baryon one. We also discuss the $N(1535)$ in the chiral doublet picture.

\end{abstract}

\ccode{PACS Nos.: 14.20.Gk,11.30.Rd,12.39.Fe,11.10.Gh,13.40.Gp,21.85.+d}

\section{Introduction}	
The $N(1535)$ is the lowest lying nucleon resonance with $J^{p}=\frac{1}{2}^{-}$, sitting just 50 MeV above the $\eta N$ threshold and having a large coupling to $\eta N$ in its decaying channel.  The $N(1535)$ has been studied in two aspects;  one is based on meson-baryon dynamics and the other on the symmetry of QCD. 
Since the $N(1535)$ is a decaying resonance, the $N(1535)$ may have large meson-baryon components. So the dynamics of mesons and baryons is important to understand the structure of the $N(1535)$. 
In a more fundamental viewpoint of chiral symmetry of QCD, the $N(1535)$ can be 
a candidate of the chiral partner of the ground state nucleon.
The chiral partners form a set of particles which transform each other under the linear realization of the chiral transformation, like $(\rho, a_{1})$ and $(\sigma, \pi)$.  
It is an interesting question how to make these two aspects consistent with each other. 




\section{Dynamical approach}

To describe baryonic resonances based on chiral dynamics, we have two ways;
1) We introduce baryonic fields explicitly in chiral effective Lagrangians as elementary particles and its dynamical properties, such as decay, are described by the coupling to the ground state mesons and baryons.  
The typical examples are the decuplet baryons. 2)  We generate resonances dynamically 
in meson-baryon scatterings with interactions described by the chiral effective theory.
The $\Lambda(1405)$ is one of the successful examples. 
In the second case, the baryonic resonances are considered to be quasi-bound states of a meson and a baryon, while in the first case the origin of the resonances is other than hadronic dynamics, such as a genuine quark state.

The chiral unitary model~(ChUM) is a powerful theoretical tool to investigate baryonic resonances in the meson and baryon dynamics.\cite{Kaiser:1995eg,Oller:2000fj} Properties of $N(1535)$ is reproduced well by the ChUM.\cite{ChUM:N1535,Lutz:2001yb,aNstar} Questions are whether the $N(1535)$ is composed by only the meson-baryon components and whether we have no room in the $N(1535)$ for the chiral partner of the nucleon.
To answer the questions, we revisit the ChUM.\cite{Hyodo:2008aa}

In the ChUM, scattering amplitudes $T$ are obtained by solving the scattering equation
$
   T = V + VGT = (V^{-1} - G)^{-1}
$,
based on the $N/D$ method,
in which the on-shell factorization leads to algebraic solution.\cite{Oller:2000fj}
$V$ is a real function describing meson-baryon interactions and given by the chiral effective theory, and $G$ restores unitarity of the scattering amplitudes and is obtained by 
the once-subtracted dispersion relation of the two body phase space with a subtraction constant $a$.

In conventional ChUM for $s$-wave baryon resonances, 
the interaction $V$ is given at first by, for example, 
the Weinberg-Tomozawa (WT) interaction, $V_{WT}(W) = -C (W-M) /(2f^2)$ with the baryon mass $M$ and the cm energy $W$, and the subtraction constants are fitted phenomenologically to reproduce the scattering amplitude:
\begin{equation}
T(W) = [V^{-1}_{WT}(W)-G(W;a_{\rm pheno})]^{-1} . \label{pheno}
\end{equation}
It is important, however,  to note that, in the renormalization point of view, once the 
scattering amplitude $T$ is fixed, change of the renormalization parameter in $G$ should be absorbed to the interaction $V$. This implies that one cannot determine $V$ and $G$ {\it a priori} and 
the equivalent scattering amplitudes can be expressed by different sets of $V$ and $G$ depending on the renormalization scheme labeled by $a$:\cite{Hyodo:2008aa}
\begin{equation}
T(W) = [V^{-1}(W;a)-G(W;a)]^{-1}. \label{general}
\end{equation}

Here we propose a renormalization scheme, 
in which the loop function $G$ is suited to the meson-baryon picture of resonances,
and see which interaction
kernel is obtained in this scheme. There the subtraction constant is fixed so as to exclude any states below the threshold and to be consistent with chiral expansion, which
can be achieved by $G(M;a_{\rm natural}) = 0$,\footnote{The details are given in Ref.~\refcite{Hyodo:2008aa}. This condition was proposed in different contexts in Refs.~\refcite{Lutz:2001yb,Igi:1998gn}.} and the interaction $V(W;a_{\rm natural})$ is left to be fixed. We call this natural renormalization scheme. 
If we obtain a good phenomenological description of the scattering amplitude with an appropriate $a_{\rm pheno}$ as in Eq.~(\ref{pheno}),
we can obtain $V(W;a_{\rm natural})$ by equating Eqs.~(\ref{pheno}) and (\ref{general}). After some algebra, 
we find in a single channel case\footnote{Generalization to the coupled channel is straightforward and discussed in Ref.~\refcite{Hyodo:2008aa}} 
\begin{equation}
V(W;a_{\rm natural})= V_{WT}(W) + \frac{C}{2f^2}\frac{(W-M)^2}{W-M_{\rm eff}}
\label{Vnatural}
\end{equation}
with $M_{\rm eff} \equiv M - \frac{2f^2}{C \Delta a}$ and $\Delta a \equiv  G(W;a_{\rm natural})-G(W;a_{\rm pheno}) = a_{\rm natural}-a_{\rm pheno}$ from the linear dependence of $a$ in $G$.
As seen in Eq.~(\ref{Vnatural}),  the interaction kernel in the natural renormalization scheme is expressed by the WT term and a pole term with the mass $M_{\rm eff}$ depending on the difference of two renormalization schemes $\Delta a$.

The relevance of the pole term depends on the value of $M_{\rm eff}$. Using 
the values of $a_{\rm pheno}$
obtained in the coupled channel \cite{aNstar} and taking a natural renormalization scheme $G(M_{N};a_{\rm natural}) = 0$ for all channels, 
we find the effective mass $M_{\rm eff} \approx 1700\pm 40i$ MeV,
which quantitatively has moderate dependence on the values of $a_{\rm pheno}$ and choice of the natural renormalization condition in the coupled channel.
Qualitatively, however, the pole mass appears in the relevant energy scale of the $N(1535)$ physics. Thus, this pole can be a source of the $N(1535)$ having some components other than dynamically generated one by meson and baryon, which can be interpreted as a genuine quark component. For the $\Lambda(1405)$ in contrast, the effective mass is found $M_{\rm eff}\approx17$ GeV, which is irrelevant for the $\Lambda(1405)$.
We consider that this is the reason that the meson-baryon picture works well for~the~$\Lambda(1405)$.

Recently transition form factors of the $N(1535)$ have been discussed 
in the meson-baryon picture.\cite{Jido:2007sm} 
There the $N(1535)$ is described by 
the ChUM with the phenomenological renormalization
scheme, and the electromagnetic transitions $\gamma^{*}N \rightarrow N(1535)$ 
are studied.  
This model fairly reproduces the observed helicity amplitudes, 
especially the neutron-proton ratio $A_{1/2}^{n}/A_{1/2}^{p}$ in good agreement
with experiments. Although this model implicitly has the quark-originated  pole for the $N(1535)$ as discussed above, direct photon couplings to the
quark components have not been considered in this model. Thus, the success of this model implies that the
meson-baryon components of the $N(1535)$ are essential
for the structure of the $N(1535)$ when probed by a low-energy virtual photon. 

\section{Symmetry aspect}
It is interesting to explore chiral partner of the nucleon in the viewpoint of fundamental symmetry of QCD. 
There are two types of linear realizations of chiral symmetry in the baryonic sector;
so-called the naive and mirror realizations, which give distinct phenomenological consequences as discussed in Ref.~\refcite{ChiralDoublet}. In the mirror realization, particularly,  the nucleon $N$ forms a chiral doublet together with a negative parity nucleon resonance $N^{*}$ and has a finite mass when chiral symmetry is restored. In both realizations, the mass difference between $N$ and $N^{*}$ is expressed by the chiral condensate and $N$ and $N^{*}$ degenerate in the chiral restoration limit. 

Assuming the chiral partner of the nucleon to be $N(1535)$, we discuss an interesting application of the chiral doublet picture to the $N(1535)$ in nuclear medium and eta mesic nuclei.\cite{EtaNuclei,Jido:2008ng} 
Recalling the fact that the $N(1535)$ is sitting just 50 MeV above the $\eta N$ threshold in the vacuum and has strong coupling to $\eta N$, we expect that the optical potential of the eta meson in nucleus has strong dependence on the in-medium modification of the $N(1535)$. Especially, if we take the chiral doublet picture and assume the partial restoration of chiral symmetry in nuclear medium, the mass difference of $N$ and $N(1535)$ is reduced as the nuclear density increases. Consequently level crossing of the eta and $N(1535)$-hole modes takes place in the nuclear medium, and the eta optical potential has strong energy and density dependence.\cite{EtaNuclei,Jido:2008ng} A recent work has found that the level crossing gives strong influence on formation spectra of the eta mesic nuclei.\cite{Jido:2008ng}

\section{Conclusion}
We have discussed the structure of the $N(1535)$ resonance both in the dynamical and symmetry aspects. The chiral unitary model gives a good phenomenological description of the structure of the $N(1535)$, while it implicitly has a source of the $N(1535)$ resonance originated by the genuine quark component, which gives us room for the chiral partner of the nucleon. The chiral doublet picture can be investigated by formation
spectra of eta mesic nuclei.

\section*{Acknowledgments}
D.J.\ acknowledges M.\ D\"oring, E.\ Oset, E.E.\ Kolomeitsev, H.\ Nagahiro and S.\ Hirenzaki for discussion. 
The work of D.J. was partially supported by the
Grant for Scientific Research (No.~18042001).
This work is part of Yukawa International Program for Quark-Hadron Sciences.

\end{document}